# On-chip ultra-compact hexagonal boron nitride topological ring-resonator in visible region

Min Wu,[1,2] Yibiao Yang,[1,2,*] Hongming Fei,[1,2,*] Han Lin,[3,*] Xiaodan Zhao,[1,2] Lijuan Kang[1,2]

[1]*College of Physics and Optoelectronics, Taiyuan University of Technology, Taiyuan 030024, China*
[2]*Key Laboratory of Advanced Transducers and Intelligent Control System, Ministry of Education, Taiyuan University of Technology, Taiyuan 030024, China*
[3]*Centre for Translational Atomaterials, Faculty of Science, Engineering and Technology, Swinburne University of Technology, Hawthorn, Victoria 3122, Australia*
* *Corresponding authors: yangyibiao_tyut@sohu.com; feihongming@tyut.edu.cn; hanlin@swin.edu.au*

**Abstract**: Ultra-compact topological ring-resonators with chirality are important devices for quantum optics. However, there are limited demonstrations of chiral resonators, especially in the visible region. We proposed a topological photonic ring-resonator based on hexagonal boron nitride (hBN) valley photonic crystal (VPC). The spin-valley locking effect in VPC allows achieving robust unidirectional transmission of edge states in the visible region (600 nm-650 nm). As a result, a high quality factor (679.3) with a free spectral range of 15.2 nm in the visible region can be achieved in a hBN all-pass filter with a compact size (2.1 μm*1.8 μm). In addition, we investigated the transmission properties of hBN ring-resonators with different shapes and combinations, confirming the flexibility of designing topological ring-resonators based on this principle. This design can be readily integrated with quantum photonic chips for broad applications.

1. Introduction

On-chip integration of quantum photonic devices offers a very promising way for realizing scalable quantum devices. The ideal quantum photonic chips have several advantages, such as small size, allowing the propagation of chiral light sources, room temperature operation and precise light field control. Among them, the propagation of chiral light sources in the quantum photonic chips puts forward higher requirements for the design and fabrication of devices. It is possible to couple a resonator to a quantum emitter or a chiral nanophotonic element, which provides a new approach to manipulating light-matter interactions [1]. All-pass filters (APFs) made of micro-ring-resonators are one of the most important resonator designs in quantum photonic chips, which can individually remove the desired channels without disturbing the passband [2]. However, APFs generally use silicon ring structures with large footprint, which are unsuitable for on-chip integration, and the surface roughness (due to the manufacturing process) of the resonator leads to backscattering of light, which reduces the transmission efficiency. As a result, the coexistence of the forward-propagating and the back-propagating modes in the resonator results in resonance splitting and a distorted Lorentzian-shaped spectrum [3]. Most importantly, conventional ring resonators cannot achieve chiral properties.

Recently, topological waveguides have emerged as a new type of photonic device that enables robust unidirectional light propagation on chips [4-9]. Due to the bulk-boundary correspondence[10], topological protected edge states at the interface between two photonic topological insulators have spin-dependent chirality. This intrinsic chirality of edge states makes them widely used in integrated quantum photonics [4-6]. Among different types of topological photonic crystals (TPCs), valley photonic crystal (VPC) has the advantage of simple design and large bandwidth. More importantly, VPCs can be designed based on conventional dielectric materials (e.g., silicon), which can be experimentally fabricated using current complementary metal oxide semiconductor (CMOS) nanofabrication techniques. As a result, it can further push the working bandwidth to shorter wavelength region, such as telecommunication [7, 11, 12] and visible region[13]. In addition, different applications based on VPC structures have been demonstrated recently, such as topological all-optical logic gates [14], robust delay lines [15], topological lasers [16], topological unidirectional transmission devices [7-9], topological resonators [17-19], and so on. Chiral ring-resonators are important functional devices for quantum optics [1], the performance of which can be judged by the coupling efficiency, the quality factor (Q factor) and free spectral range (FSR). It has been demonstrated a topological ring-resonator based on silicon VPC structure can have high Q factor (up to 1000) and high coupling efficiency (18 dB) at telecommunication wavelength (1550 nm). However, topological resonators in the visible region have not been demonstrated. It is desired to find new type of dielectric material for designing ultra-compact topological ring-resonators operating in the visible region.

Two-dimensional hexagonal boron nitride (2D hBN) [20-24], also known as "white graphene", due to its high mechanical strength, high thermal stability and excellent chemical properties. It has demonstrated broad applications in the field of optically stable ultra-bright quantum single-photon light sources[20, 21]. Recently, it is demonstrated that a experimentally fabricated 2D hBN PC cavity [23] can be used as a quantum single-photon light source in visible range with ultra-high brightness at room temperature, confirming the feasibility of experimental preparation of hBN PCs working at visible to near-infrared wavelengths. Therefore, it is meaningful to design VPC devices based on 2D hBN material, which can directly couple and process the on-chip single photon source from hBN for quantum computing[13, 22, 23]. One of the scenarios is to enhance the single photon emission by coupling the quantum emitter into a topological ring-resonator[19], which desired a design of high-performance chiral ring-resonator based on 2D hBN material.

Here we theoretically proposed a topological chiral photonic ring-resonator working in visible region based on hBN material. The ring-resonator is designed using VPC structure. The spin-valley locking effect allows supporting two counter propagating edge modes with opposite polarizations. As a result, the emitter emits preferably into one of the edge modes depending upon its spin status. The edge states according to different boundaries are optimized to ensure high propagation efficiency of the ring-resonator. The length of the ring-resonators is tuned to control the number and positions of the peaks. As a result, a high quality factor of 679.3 with a free spectral range of 15.2 nm in the visible region are achieved with an ultra-compact footprint of 2.1 μm*1.8 μm. We further present different combinations and shapes of ring-resonators to show the flexibility of the design. The designed ring-resonators can be fabricated with current CMOS nanofabrication process, thus will find broad applications in quantum photonics chips.

2. Design of the topological ring-resonator

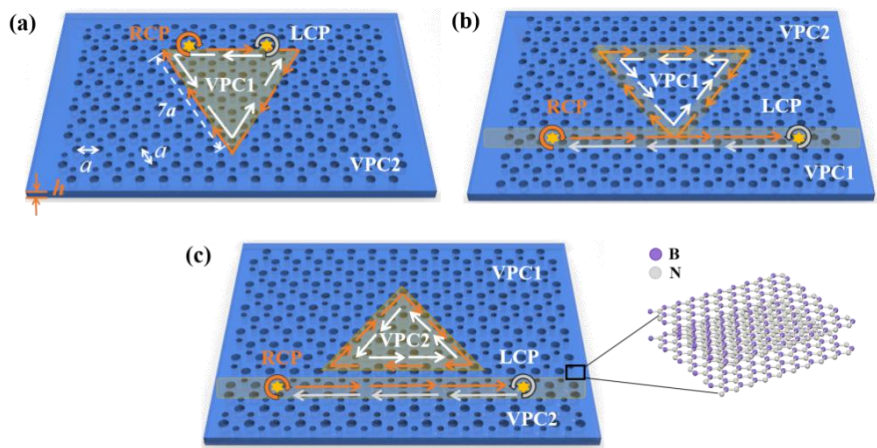

Fig. 1 (a) Three-dimensional (3D) schematic of a triangular shape topological ring-resonator based on hBN VPC structure. The propagation directions of the two counter propagating edge states (RCP/LCP) are indicated by orange/white arrows; (b) Schematic diagram of the topological ring-resonator coupled with the topological straight waveguide (indicating by the orange shadow). The light sources (RCP/LCP light) are located in the straight waveguide indicating the orange/white arrows. (c) Schematic diagram of a topological ring resonator coupled with the topological straight waveguide which is parallel to a side of the triangle. Inset: molecular structure of 2D hBN material.

We design a topological triangular ring resonator based a 2D hBN VPC structure as shown in Fig. 1(a), which is composed of two honeycomb lattice photonic crystal (PC) with circular holes embedded in a free-standing hBN plate, named VPC1 and VPC2. The arrows indicate the light propagation direction of left-handed circularly polarized (LCP) and right-handed circularly polarized (RCP) light in the ring-resonator. Both VPC lattices contain a set of large and small air holes in a unit cell for breaking the spatial inversion symmetry of the honeycomb lattice structure. The values of the Berry curvature and topologically invariant valley Chern number $C_V$ corresponding to the two VPCs were calculated, which are $C_V=-1$ for VPC1 and $C_V=1$ for VPC2 [25, 26], respectively. The lattice constant is $a = 230$ nm for both VPCs. The radii of the large and the small circle are $r1 = 60$ nm and $r2 = 30$ nm, respectively. The thickness of the freestanding hBN plate is $h = 220$ nm. As 2D-hBN has a layered structure, and the overall thickness depends on the number of layers. Furthermore, the layered structure leads to anisotropic refractive indices ($n_x=n_y \neq n_z$, where $n_x$, $n_y$ are the in-plane refractive indices and $n_z$ is the out-of-plane refractive index) [23, 27], which is considered in the design. The hBN material has low absorption in the visible region, which allows efficient unidirectional transmission of visible light. To explore the coupling between the emitter and topological edge state modes, the quantum emitter is embedded on the ring-resonator. In addition, in order to construct an APF, we combined ring-resonators with straight topological waveguides, which may have two different combinations as shown in Figs. 1(b) and (c). In these cases, the quantum emitters are placed on the straight waveguides. The light paths of the light waves are indicated by the orange and white arrows.

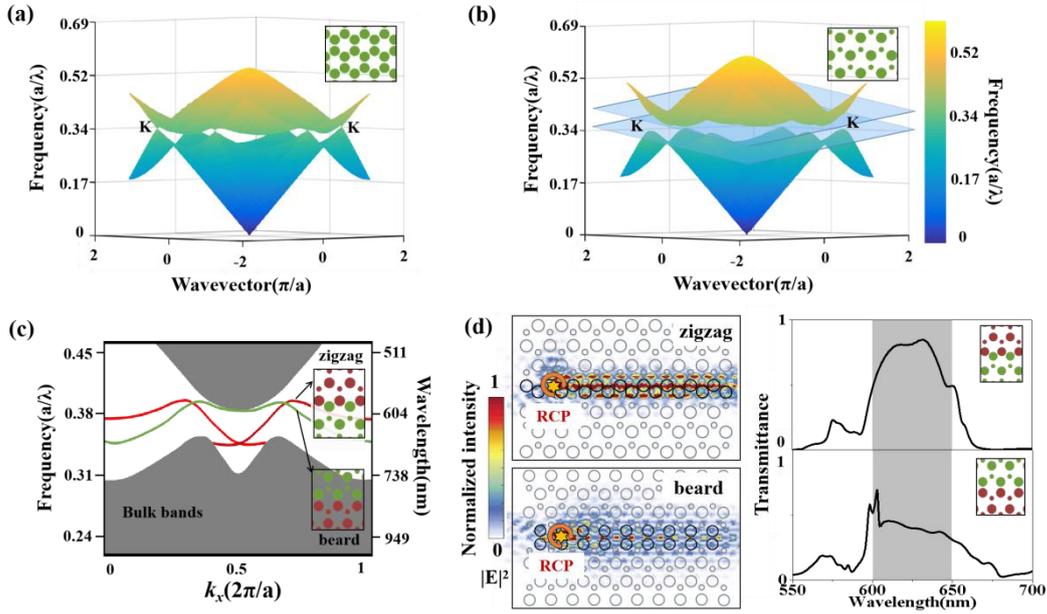

Fig. 2 (a) The 3D band diagram of the honeycomb lattice structure, the Dirac point K is marked in the figure; (b) The 3D band structure of VPC1 and VPC2; (c) Dispersion diagram of spin-valley locked edge states at the interface between VPC1 and VPC2; (d) Electric field intensity distribution and spectra of different types (zigzag and beard) of straight waveguides.

In order to realize the spin-valley locking effect, we first build a honeycomb PC structure with $C_{6v}$ symmetry ($r1=r2=45$ nm) to achieve a Dirac point in the band structure (Fig. 2(a)). By enlarging the radii of a set of air holes and reducing the other, the $C_{6v}$ rotation symmetry can be reduced to $C_{3v}$, which opens a topological photonic bandgap at the K point to form VPC1 and VPC2 (Fig. 2(b)). The bandgap is in the wavelength range of 0.354-0.38 a/λ (600 nm-650 nm), indicated by the gap between the two blue transparent plates. There are two possible combinations to construct a boundary by using VPC1 and VPC2, namely the zigzag and beard geometries, which support different topological edge states (Fig. 2(c)). Due to the spin-valley locking effect, different polarizations will be locked to different valley (K or K' valley). As shown in Fig. 2(d) by the electric field intensity distribution and the transmittance spectra, the zigzag boundary has much higher forward transmittance than the beard boundary within the working bandwidth of 600 nm-650 nm, which is approaching 1 and much higher than unidirectional transmission device based on conventional PC structures [28-37]. Therefore, in this paper, we will mainly focus on the ring-resonators based on the zigzag boundary.

3. Tuning the length of topological ring-resonator

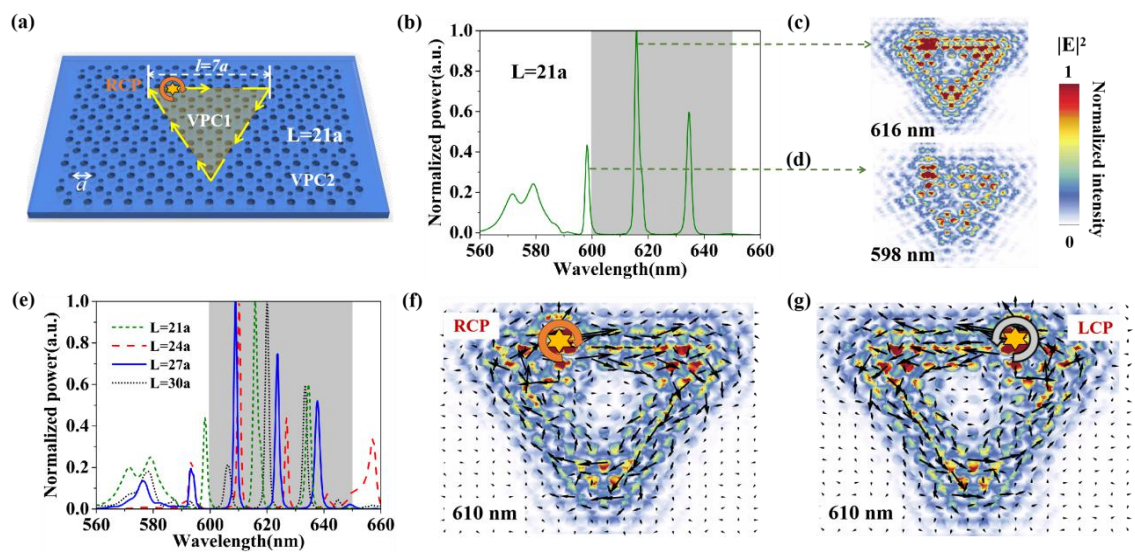

Fig. 3 (a) Schematic diagram of the topological ring-resonator. The quantum emitter is marked in the figure. (b) The normalized optical power spectra of the ring-resonator with a total length of L=21$a$, corresponding to 7$a$ on each side. The shaded gray area corresponds to the topological bandgap, and the peaks within the gray area correspond to the resonance modes. (c, d) The normalized electric field intensity distributions of the topological resonance mode (616 nm) and bulk state mode (598 nm), respectively. (e) Normalized power spectra of topological ring-resonators with different lengths. Simulated Poynting vector energy flow diagrams within the topological ring-resonator of RCP(f) and LCP(g) light, respectively.

Different from conventional waveguides, the topological edge states can achieve robust unidirectional transmission at sharp bends. Therefore, we construct a triangular ring-resonator (Fig. 3(a)). The properties of the ring-resonator are studied by using the 3D finite difference time domain (FDTD) method. Here the total length of the ring (L) is defined as a variable to control the number and the positions of the resonance peaks. The structure design of a ring-resonator with L=21$a$ (4.83 μm) is shown in Fig. 3(a). The power spectrum is shown in Fig. 3(b), which shows 3 peaks in total. Since the working bandwidth of the topological edge states is in the range of 600 nm-650 nm, the peak 598 nm belongs to the trivial bulk states, thus shows relatively low power, which is also confirmed by the messy electric field intensity distribution shown in Fig. 3(d). In comparison, the other two peaks at 616 nm and 635 nm belong to the topological edge states. The free spectral range (FSR) is 19 nm. The Q factors are 341.8 and 352.3 at the peaks of 616 nm and 635 nm, respectively. From the field intensity distribution (Fig. 3(c)) one can see that the electric field is well confined within the ring-resonator without strong scattering even at the sharp corners due to the spin-valley locking effect. As expected, the number of peaks can be increased by increasing the length of the ring resonator as demonstrated in Fig. 3(e), which results in a decrease of FSR. In addition, the Q factor can be increased accordingly. One example is L=27$a$ (6.21 μm), which shows 3 peaks in the working bandwidth, separated by FSR of 14.4 nm. The Q factor is 507 at the peak of 609 nm. In order to demonstrate the spin-dependent propagation behavior of light in the structure, we plot the Poynting vector distributions in a ring-resonator (L=27$a$) in Figs. 3(f) and (g). The RCP and LCP light propagate clockwise and counterclockwise, respectively. The schematic diagrams and electric field intensity distributions of the ring-resonators with different lengths are shown in the Supplementary Section 1.

4. Design of all pass filters

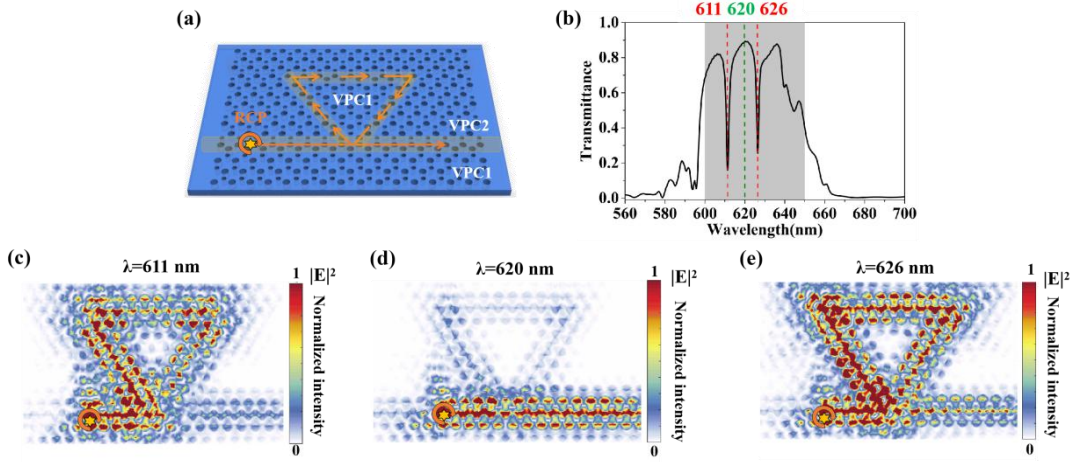

Fig. 4 (a) Schematic diagram of the coupling between a ring-resonator and a straight topological waveguide to form an APF. The orange arrow represents the propagation direction of the RCP light. (b) Transmission spectra of the APF. The electric field intensity distribution of the APF at the peaks of 611 nm (c), 620 nm (d) and 626 nm (e).

We further couple the ring-resonator with a straight topological waveguide to form an all-pass filter (APF) (Fig. 4(a)), which is one of the most commonly used functional devices based on ring-resonator. In order to maximize the working bandwidth of the APF, we choose the straight waveguide with the same zigzag boundary as the ring-resonator. In this design, the straight waveguide is connected to a corner of the ring-resonator. Here is the length of the ring-resonator is L=27$a$. As expected, the APF shows high transmittance (>0.9) within the working bandwidth, except for the dips coupled to the ring-resonator (Fig. 4(b)). The transmittance spectrum is shown that the two dips are separated by a FSR of 15.2 nm, and the Q factor is 679.3 at the peak of 611 nm. To show the dips come from the efficient coupling to the resonance modes of the ring-resonator, the electric field distributions at different wavelengths are plotted in Figs. 4(c)-(e). One can see that the resonance modes are circulating within the ring-resonator (Figs. 4(c) and (e)). In comparison, at the wavelength of the resonance dips, the light pass through the straight waveguide without coupling into the ring-resonator (Fig. 4(d)). The effects of the distance between the triangular ring-resonator and the straight waveguide are studied in the Supplementary Section 2. Assuming lossless balance, i.e., considering topological properties and a 50:50 beam splitting process at the branch (the position of the corner of the ring-resonator touches the straight waveguide) that is structurally symmetric, the final transmission of the APF can be written as [17]:

$$T = \frac{0.5 + a_l^2 - \sqrt{2} a_l \cos\delta}{1 + 0.5 a_l^2 - \sqrt{2} a_l \cos\delta} \quad (1)$$

where $a_l$ and $\delta = 6\pi n_{eff} l/\lambda$ are the round-trip loss coefficient and phase shift, respectively. $n_{eff}$ is the effective refractive index of the propagating mode, $l$ is the side length of the triangular ring, and $\lambda$ is the working wavelength. When the phase shift $\delta$ is a multiple of $2\pi$, the light resonates and shows a dip in the transmittance spectrum. Note that the resonance mode is topologically protected against the path direction of the ring-resonator combined with a straight waveguide, which has a counterclockwise orientation and is coupled to the straight waveguide by an evanescent field [2, 17].

Considering the experimental feasibility of the designed hBN APF, we further study the transmittance spectra of the hBN APF with different experimentally measured in-plane dispersive refractive indices [27, 38, 39] (Supplementary Section 3). The APFs with different refractive indices show high filtering performance. As a result, high refractive index is preference to achieve a high Q-factor and strong suppression at the dips.

## 5. Designs of the ring-resonators with different shapes

One potential way to tune the performance of the APF is to combine ring-resonators with different lengths. Due to the different positions of the peaks from different ring-resonators, it is expected to show multiple peaks within a short interval. Two examples are shown in Figs. 5(a) and (c), which have the combinations of ring-resonators. In the case of combining ring-resonators with L=27$a$ and L=18$a$, due to the overlapped the peak positions at 606 nm and 625 nm, the suppression is significantly increased. The transmittance is approaching 0, which improve the performance of the APF device. In comparison, when the ring-resonator with L=27$a$ is combined with the one with L=21$a$, there are more dips shown in the working bandwidth. Other combinations of the ring-resonators with the same length (L=27$a$) are shown in the Supplementary Section 4. Therefore, the combination of different ring-resonators (the number and the lengths can be variables) offers a flexible method to design APF.

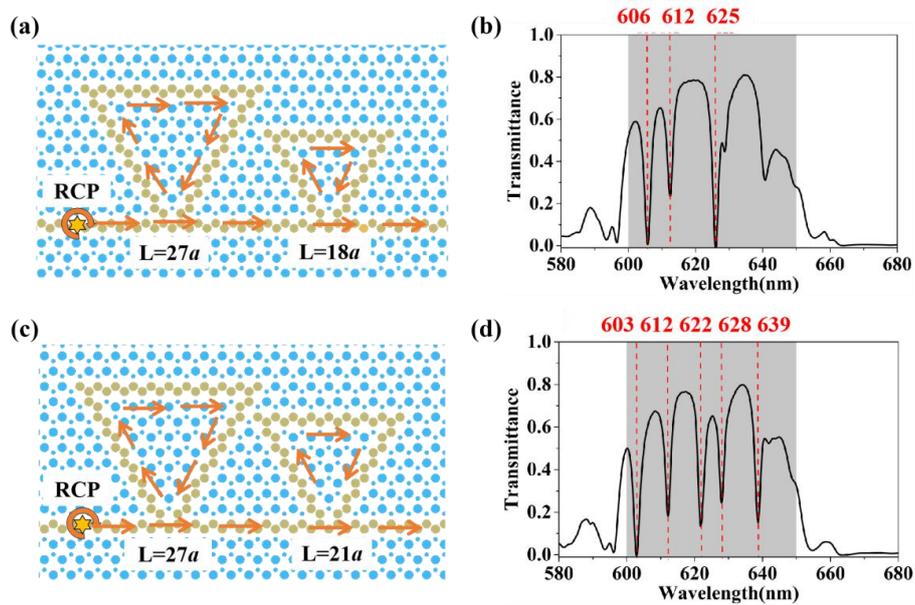

Fig. 5 Schematic diagrams and transmittance spectra of the APFs composed of double topological ring-resonators and a straight topological waveguide. (a) The lengths of the ring-resonators are L=27$a$ and L=18$a$, respectively; (b) The corresponding transmittance spectrum. (c) The lengths of the ring-resonators are L=27$a$ and L=21$a$, respectively; (d) The corresponding transmittance spectrum. The shaded area is the bandwidth of the topological edge states.

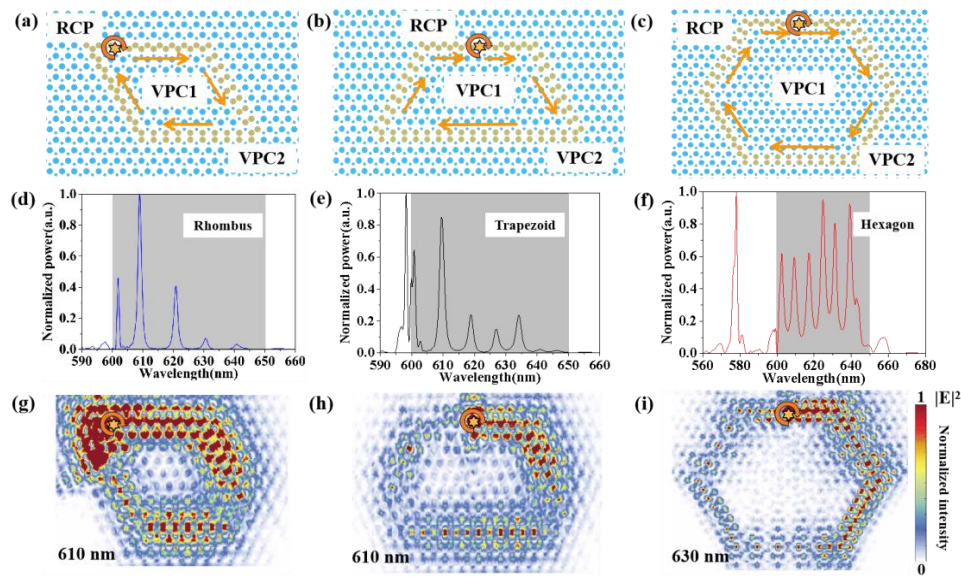

Fig. 6 Ring resonators with different shapes of zigzag and beard straight waveguides. The edge state is VPC1 (inside) and VPC2 (outside). The yellow column area is the location of the edge state. The normalized power distributions of the resonators are simulated, and the in-plane electric field strength distribution for one of the resonance modes: (a, d, g) rhombus (b, e, h) isosceles trapezoid (c, f, i) hexagonal.

In addition, since the propagation efficiency of topological edge states is hardly affected by the sharp corners, for example, 60° and 120° corners, the shapes of the ring-resonators can be used as another tunning parameter. Here the three different shaped ring-resonators are designed, namely the rhombus, trapezoid and hexagonal as shown in Fig. 6. It is found that all the three types of ring-resonators can show sharp peaks in the working bandwidth (Figs. 6(d)-(f)). In addition, from the electric field intensity distributions shown in Figs. 6(g)-(i), all the ring-resonators can strongly confine the electric field inside the perimeter of the ring-resonators.

6. Conclusion

In conclusion, we have designed topological ring-resonators based on the hBN VPC structure working in the visible wavelength region. By combining the ring-resonators with a straight topological waveguide, we further design functional APFs. The spin-valley locking effect allows achieving a high Q-factor (up to 679.3) at the 611 nm due to the robust unidirectional propagation. The designed hBN ring-resonators have ultra-compact footprints and can be fabricated using current nanofabrication techniques (The proposed fabrication method of hBN topological ring-resonators and APFs is mentioned in the Supplementary Section 5), thus, it will find broad applications in quantum chips. In particular, the chiral property of the ring-resonator allows the devices to work with on-chip quantum emitters of different spin states.